# Techniques Enabling Generator Refactoring

Holger Krahn and Bernhard Rumpe

Institute for Software Systems Engineering
Technische Universität Braunschweig, Braunschweig, Germany
http://www.sse.cs.tu-bs.de

**Abstract.** This paper presents our approach to use refactoring techniques together with code generation. Refactoring is particularly useful if not only the generated classes but also the generator itself can be adapted in an automatic fashion. We have developed a simple demonstration prototype to illustrate this. The demonstration is based on a special technique where the template for the code generation is defined as compilable source code. The directives to fill out this template prototype to the actual classes are embedded in the source as comments.

## 1 Overview

Code generation avoids repetitive programming tasks and helps to improve code quality. When code generators are used in agile projects, one problem occurs: The hand-coded source code is frequently changed using existing refactoring [2] tools. But either the code generation is not repeatable (one shot only) or the equivalent changes in the code generator have to be applied manually.

The key idea of this abstract is to use a template-based code generation where the template describing the target is at the same time a compilable class. This should enable existing tools to refactor the generator and not the generated code only. In this way a round-trip approach and its potential problems can be avoided. Figure 1 shows an overview of the applied tool chain.

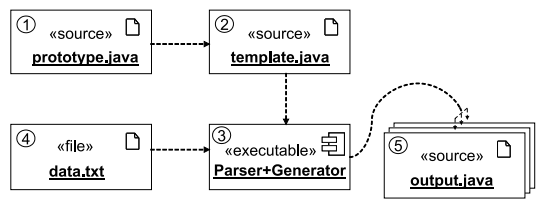

**Fig. 1.** Overview of the tool chain

Five steps have to be applied to use the code generator, which are illustrated by a simple example:

1. A prototype for the generated code is programmed manually, an example can be found in figure 2.
2. The variation points (template holes) of the class are identified and special comments are added directly before the tokens to be replaced. For simplicity of our demonstration, the source is tokenized using spaces as separators. For an example see figure 3.

3. A parser for the input data can be written e.g. using a parser generator. The generator is the same for every template class.
4. The data for the variation points has to be specified. An example input for the parser from step 3 can be seen in figure 4.
5. The Parser+Generator component combines the data and the template to the generated classes. An example output can be found in figure 5.

```
class A {
  public String toString() {
    Printer.write("A");
  }
}
```

**Fig. 2.** Example prototype class

```
class /*C %name% */ A {
  public String toString() {
    Printer.write( /*C " %name% " */ "A" );
  }
}
```

**Fig. 3.** Example template class

```
name = Generated;
name = Bar;
name = Foo;
...
```

**Fig. 4.** Example for repeated use

```
class Generated {
  public String toString() {
    Printer.write("Generated");
  }
}
```

**Fig. 5.** Example generated class

In the given example, refactorings like "rename method" which might rename `Printer.write(...)` to `Printer.print(...)` can be applied easily.

## 2 Related Work

Various ways of code generation are already published. For a survey of the most common approaches see for example [3]. Most similar to our approach are template-based code generators like Velocity [1]. All these approaches have in common that they use a separate template language which results in files that cannot be directly compiled by a conventional compiler and therefore not directly be refactored.

## 3 Conclusion

We implemented a tool chain for demonstration purposes which supports - in addition to the explained mechanisms above - iterative adaptation within a template class, without affecting the generator as such. Based on this tool, we are now exploring its advantages and limitations. In particular it is unclear by now, where such an approach breaks down at all, because it is also possible to add control structures in form of `/*C forall ...*/` or `/*C if ... */` etc. As further work, we try to overcome some of the limitations, such as an improvement of the tokenizing form, and use this approach in larger case studies.